\documentclass[twocolumn]{openjournal}

\usepackage{orcidlink} 
\usepackage[T1]{fontenc}
\usepackage{ae,aecompl}

\usepackage{graphics,epsf}
\usepackage{amsmath}                
\usepackage{amsfonts}               
\usepackage{amssymb}                
\usepackage{epsfig}                 
\usepackage{appendix}
\usepackage{graphicx}
\usepackage{float}
\usepackage{color}
\usepackage{multirow}
\usepackage{colortbl}
\usepackage[para,online,flushleft]{threeparttable}

\hypersetup{citecolor=blue, 
            linkcolor=red, 
            menucolor=blue, 
            urlcolor=blue}  

\newcommand{\cm}{{~\rm cm}}
\newcommand{\m}{{~\rm m}}
\newcommand{\km}{{~\rm km}}
\newcommand{\s}{{~\rm s}}
\newcommand{\h}{{~\rm h}}

\newcommand{\g}{{~\rm g}}
\newcommand{\G}{{~\rm G}}
\newcommand{\K}{{~\rm K}}
\newcommand{\erg}{{~\rm erg}}
\newcommand{\yr}{{~\rm yr}}

\newcommand{\days}{{~\rm days}}

\begin{document}

\title{Jet-shaped filamentary ejecta in common envelope evolution}

\author{Ron Schreier}
\affiliation{Department of Physics, Technion Israel Institute of Technology, Haifa, 3200003, Israel; \\ 	
ronsr@physics.technion.ac.il, shlomi.hillel@gmail.com, soker@physics.technion.ac.il}

\author{Shlomi Hillel}
\affiliation{Department of Physics, Technion Israel Institute of Technology, Haifa, 3200003, Israel; \\ 	
ronsr@physics.technion.ac.il, shlomi.hillel@gmail.com, soker@physics.technion.ac.il}

\author{Noam Soker\,\orcidlink{0000-0003-0375-8987}} 
\affiliation{Department of Physics, Technion Israel Institute of Technology, Haifa, 3200003, Israel; \\ 	
ronsr@physics.technion.ac.il, shlomi.hillel@gmail.com, soker@physics.technion.ac.il}

\begin{abstract}
We conduct three-dimensional (3D) hydrodynamical simulations of common envelope evolution (CEE) of a neutron star (NS) that launches jets as it spirals in inside the envelope of a rotating red supergiant (RSG) stellar envelope and find that Rayleigh-Taylor instabilities form filamentary ejecta. We first study the 3D RSG envelope properties before we launch the jets. Adding envelope rotation causes the RSG envelope to expand in the equatorial plane and contract along the poles, leading to non-radial oscillations that decay after two oscillation periods, like the radial oscillation of the non-rotating model. In addition, the envelope becomes convective with large vortices, as in the non-rotating case. Since RSG stars oscillate and have envelope convection, we strengthen the claim that there is no need to relax one-dimensional stellar models of cool giant stars when transporting them to 3D grids. When adding jets, the 3D simulations that include pre-set envelope rotation show that envelope rotation leads to more prominent spiral structures of the ejecta than in the non-rotating case. We map the envelope zones that are Rayleigh-Taylor unstable and conclude that this instability forms the filamentary ejecta, with and without envelope rotation. The jet-inflated high-pressure volumes around the NS accelerate the envelope, a process prone to Rayleigh-Taylor instability. 
\end{abstract}

\keywords{(stars:) binaries (including multiple): close; (stars:) supernovae: general; transients: supernovae; stars: jets} 

\section{Introduction} 
\label{sec:intro}

In a common envelope evolution (CEE), a compact object, the secondary, orbits inside the extended envelope of the primary star. The interaction of the secondary star with the envelope is not symmetrical, i.e., it has no cylindrical symmetry because of the density gradient in the envelope and the circular, elliptical, or spiraling-in orbit. Consequently, the mass that the secondary accretes has a net-specific angular momentum. If the secondary is sufficiently compact, the accretion process proceeds through an accretion disk around the secondary. As some theoretical studies (e.g., \citealt{ArmitageLivio2000, Chevalier2012, LopezCamaraetal2019, LopezCamaraetal2020MN}) showed, such accretion disks in CEE might launch jets, mainly when the secondary is a neutron star (NS) or a black hole (for a recent review see \citealt{Grichener2025}). Also, main sequence stars in CEE might also launch jets (e.g., \citealt{BlackmanLucchini2014, Soker2023MS}). 
 The disk around the NS might continuously exist from the pre-CEE phase, like from the Roche lobe overflow process (e.g., \citealt{JuarezGarciaetal2025}).  
The jets remove mass from the secondary vicinity, hence reducing mass accretion rate, and by that, operate in a negative feedback mechanism, e.g., \cite{Soker2016Rev} for an old review, and  \cite{Gricheneretal2021} and \cite{Hilleletal2022FB} for one-dimensional (1D) and 3D simulations, respectively, of this negative feedback mechanism. Alongside the negative, there is a positive feedback mechanism because the jets that the accretion disk launches carry energy that reduces the pressure around the secondary (e.g., \citealt{Shiberetal2016, Chamandyetal2018, LopezCamaraetal2019, LopezCamaraetal2020MN}). 

Most 3D simulations of the CEE do not include jets that the secondary star launches (e.g., \citealt{Passyetal2012, RickerTaam2012, Nandezetal2014, Staffetal2016MN, Kuruwitaetal2016, Ohlmannetal2016a,  Iaconietal2017b, Chamandyetal2019, LawSmithetal2020, GlanzPerets2021a, GlanzPerets2021b, GonzalezBolivar2022, Lauetal2022a, Lauetal2022b,  BermudezBustamanteetal2024, Chamandyetal2024, Gagnieretal2024, GonzalezBolivaretal2024, Landrietal2024, RosselliCalderon2024, Vatteretal2024, Vatteretal2025B}, for a small fraction of papers presenting CEE simulation; \citealt{RoepkeDeMarco2023} for a review). The small number of 3D hydrodynamical studies of the CEE and the grazing envelope evolution that includes the jets that the secondary launches have to give up other ingredients of the CEE  (e.g., \citealt{MorenoMendezetal2017, Shiber2018, ShiberSoker2018, LopezCamaraetal2019, Schreieretal2019inclined, Shiberetal2019, LopezCamaraetal2020MN, LopezCamaraetal2022, Zouetal2022, Soker2022Rev, Hilleletal2023, Schreieretal2023, Gurjareta2024eas, ShiberIaconi2024}). Unrelated to our present study are simulations that obtain collimated outflows from the distorted envelope at the final phases of the CEE (e.g., \citealt{Zouetal2020, Morenoetal2022, Ondratscheketal2022, Vatteretal2024}), as suggested many years ago by an analytical calculation \citep{Soker1992}. 

An NS or a black hole that launches jets inside the envelope, and then the core of a red supergiant (RSG) can power a very energetic event that mimics a peculiar supernova (e.g., \citealt{Sokeretal2019}); it is termed a common envelope jets supernova (CEJSN) event \citep{SokerGilkis2018}. If the NS or black hole does not reach the core and launches the powering jets only in the envelope, the event is a CEJSN impostor (\citealt{Gilkisetal2019}). CEJSN can lead to r-process nucleosynthesis inside the jets (e.g., \citealt{GrichenerSoker2019, JinSoker2024}), and the jets of CEJSN impostors with black holes might be sources of energetic neutrinos \citep{GrichenerSoker2021}; for a review see \cite{Grichener2025} and on the occurrence rate see, e.g., \cite{Grichener2023, Ungeretal2025}. 

In this study, we widen our exploration of CEJSN impostors by including pre-CEE stellar rotation in 3D simulations of the CEJSN importers with NS companions. The basic numerical procedure and numerical settings are similar to those in our earlier study \cite{Hilleletal2023}; we briefly discuss these in Section \ref{sec:Numerical}.
In Section \ref{sec:StellarModel}, we describe the properties of the rotating stellar envelope before we launch jets into the envelope, and in Section \ref{sec:Ejecta}, we describe some effects of the jets. 
We summarise the main results of the CEJSN imposter with rotation simulations in section \ref{sec:Summary}.

\section{The numerical setup}
\label{sec:Numerical}

We described our numerical setup in \cite{Hilleletal2023}. In the following, we will briefly outline the main ingredients.

The first ingredient is the non-rotating stellar model. We use a zero-age-main-sequence star of metalicity $Z=0.02$ and mass $M_{\rm 1,ZAMS}=15 M_\odot$ evolved to the RSG phase using the one-dimensional (1D) stellar evolution code \texttt{MESA} \citep{Paxtonetal2011, Paxtonetal2013, Paxtonetal2015, Paxtonetal2018, Paxtonetal2019}. Our primary focus is 3D hydrodynamical simulations using {\sc flash} \citep{Fryxelletal2000}. We transport the 1D stellar model to the 3D numerical grid when its mass and radius are $M_1=12.5 M_\odot$ and $R_{\rm RSG}=881\,R_{\odot}$, respectively.   Cells outside the stellar model are initialized with a low density of $\rho_{\rm grid,0} = 2.1 \times 10^{-13} \g \cm^{-3}$ and a temperature of $T_{\rm grid,0}= 1100 \K$, and outflow boundary conditions are used.  

The inner $20\%$ of the stellar radius is kept fixed.   This inner inert core, with a radius of $R_{\rm inert}=176 R_\odot$, saves large amounts of computational resources, as simulating inner envelope zones requires smaller time steps.   We use a stationary gravitational field computed at the initial time from the stellar model, including the fixed inner sphere. Our numerical grid is a Cartesian grid with outflow boundary conditions. The matter occupying the grid is an ideal gas ($\gamma = 5/3$), including radiation pressure.

The second ingredient is the NS orbit. We operate in a CEJSN scenario where the NS spirals in and launches jets due to mass accretion. We use a fixed spiral orbit that mimics the plunge-in phase of the CEE.

The third ingredient is the launching of jets. The orbiting NS does not affect the simulation, except for the jets launched from its location at any given moment. 
Our scheme does not alter the mass density of any computational cell, but rather changes the velocity and internal energy of the existing material near the NS. 
The power of the two opposing jets varies with the location of the NS according to the Bondi-Hoyle-Lyttleton (BHL) mass accretion rate from the unperturbed envelope reduced by a multiplication factor $\zeta \simeq 0.002 - 0.005$ (\citealt{Gricheneretal2021, Hilleletal2022FB}).
  The mass and radius of the NS are taken to be $M_{\rm NS}=1.4 M_\odot$ and $R_{\rm NS}=12 \km$, respectively.  
In this study, we take $\zeta = 7.5 \times 10^{-5}$  for three  reasons.   (1) Much stronger jets will dominate and obscure the effect of the envelope's rotation.   (2) The topic of this study is the effect of envelope rotation, so we simulate three cases with different rotation velocities. We do not scan other parameters, such as elliptical orbits and jets' power (3). As we found in an earlier paper \citep{Hilleletal2022FB}, simulations with much more powerful jets are highly demanding and will take months to simulate. Future studies should simulate other parameters for specific needs as observations of other numerical studies appear and include radiative transport, a significant process for CEE of massive stars (e.g., \citealt{Lauetal2025}).   

The energy and momentum of the jets are inserted in a jet-envelope interaction region, which is assumed to be a cylinder   with a base radius of $4 \times 10^{12} \cm$ and total height $14 \times 10^{12} \cm$ . The scheme conserves mass, momentum, and energy.

The fourth ingredient, new in our study of jets in CEE, is preset envelope rotation. We start the 3D simulation with a solid-body envelope rotation in the spherical non-rotating model (as we transferred from the 1D simulation) with an angular velocity of 
\begin{equation}
\Omega = \alpha \sqrt{G M/R^3} ,
    \label{eq:Omega}
\end{equation}
where $M$ and $R$ are the mass and stellar radius at the beginning of the simulation. In addition to the non-rotating case \citep{Hilleletal2023}, we simulate two cases with $\alpha=0.25$ and $\alpha=0.5$. 
 For $\alpha=0.5$, on the equator, the centrifugal force is one-quarter of the gravitational force. We did not simulate cases with faster rotation because, in such cases, centrifugal forces become too large to justify the assumption that the star is spherical, despite its rotation. Simulating a much faster rotation requires starting the simulation earlier and having a higher resolution near the equatorial plane. We do not yet have the required numerical resources.  

\section{The rotating three-dimensional stellar model} 
\label{sec:StellarModel}

We describe the behavior of the 3D model that we transported from the 1D model of \textsc{mesa} (section \ref{sec:Numerical}) when we add envelope rotation (we add the rotation after we transfer the model to the 3D grid). In this section, we describe the model's behavior without jets. 

We present results obtained using both regular and high resolutions. the center of the grid is at the center of the RSG stellar model.    In the regular resolution simulations, we   use a uniform grid with a cell size of $L_{\rm G}/128 = 3.90625 \times 10^{12} \cm$, where $L_G=5\times 10^{14}$   is the size of the simulation   cube  . In the high-resolution simulations, the cells have a size of $L_{\rm G}/256 = 1.95 \times 10^{12} \cm$. If we do not specify the resolution, it defaults to the regular one. Due to limited computational resources, we use only the regular resolution when including jets (Section \ref{sec:Ejecta}). 
We take the $ z$-axis along the initial envelope's angular momentum.  

In our study of the non-rotating 3D model \citep{Hilleletal2023}, we found that the envelope performs two oscillations and then decays. We argued that this is an expected behavior of RSGs, and therefore, there is no need to stabilize the model before starting the CEE simulations; just let the star oscillate. We follow the initial mass in spherical shells to reveal the oscillatory behavior. We mark each initial spherical shell with a `tracer'; all cells in a shell start at $t=0$ with ${\rm tracer}=1$. The tracer decreases as the initial shell's material mixes with gas outside the shell. The tracer is a fraction of each cell's mass originating in the shell. The tracer value is always between 0 and 1. 
The two initial shells we follow are $600 R_\odot < r < 650 R_\odot$  and $800 R_\odot < r < 850 R_\odot$; the stellar radius when we start the 3D simulation is $R_{\rm RSG}=881\,R_{\odot}$.

\begin{figure} 
\centering
\includegraphics[width=0.44\textwidth]{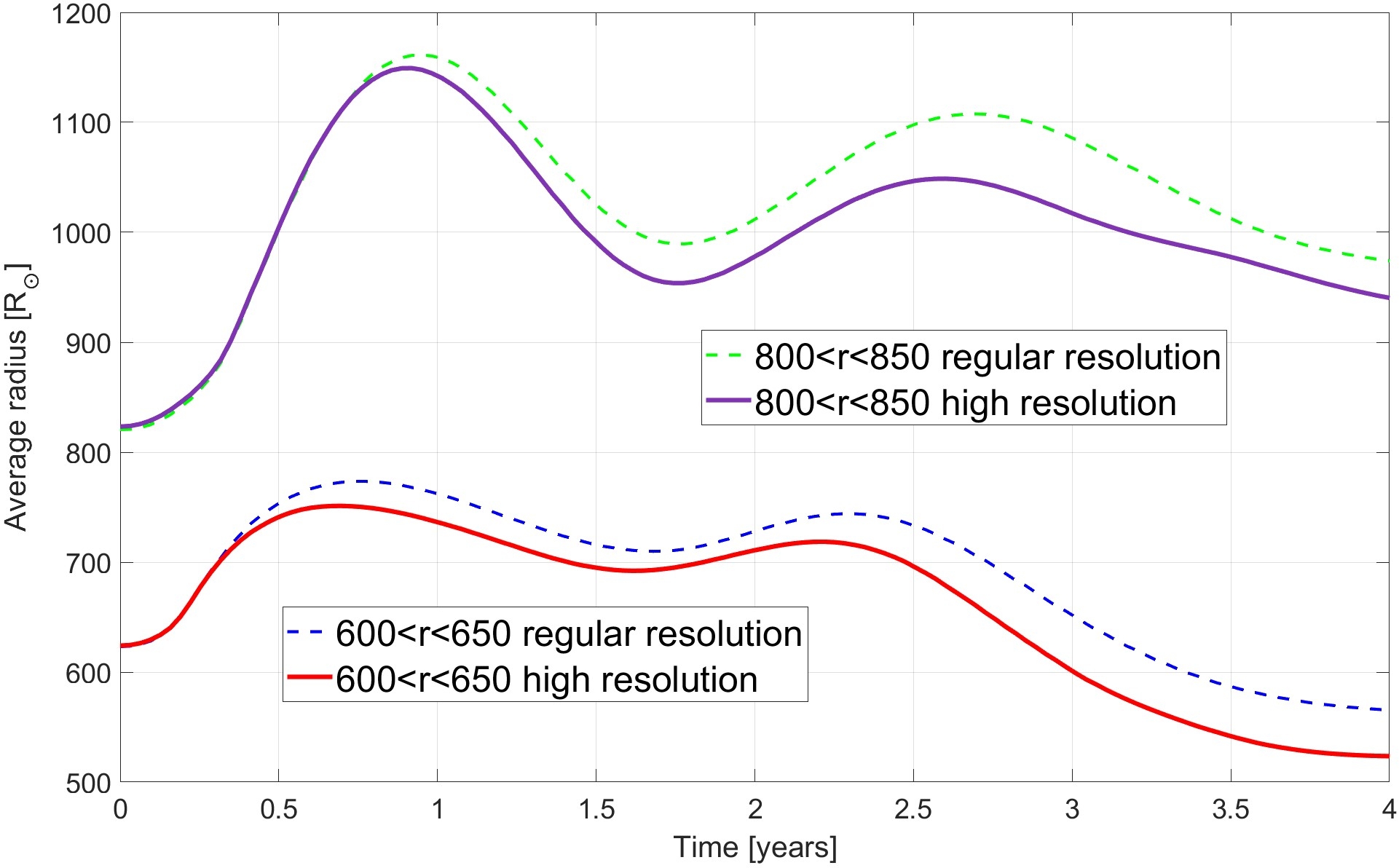}
\caption{The average radii of the tracers of two initial spherical shells $800 R_\odot < r < 850 R_\odot$ and $600 R_\odot < r < 650 R_\odot$, comparing the regular resolution (thin-dashed lines) and the high resolution (thick-solid lines) of no-jet simulations of a rotating envelope with $\alpha=0.5$. 
The differences are sufficiently small to justify simulating jets with the regular resolution.   
}
\label{fig:LowHighRes}
\end{figure}

Figure \ref{fig:LowHighRes} compares the high-resolution simulation with the regular-resolution simulation, both without jets.   This figure shows the average radii of the tracers of two initial spherical shells $800 R_\odot < r < 850 R_\odot$ and $600 R_\odot < r < 650 R_\odot$. The oscillation of the envelope for two cycles before decaying is evident. The period between the two maxima is $\simeq 1.7 \yr$. For comparison, the Keplerian orbital period on the surface of the unperturbed RSG is $P_{\rm Kep}=2.35  \yr$, and the dynamical time is $P_{\rm D}\equiv (G \bar \rho)^{-1/2} =0.76  \yr$ where $\bar \rho$ is the average density of the initial RSG model. The oscillations are of a dynamic nature (see below).   The differences between the regular and high resolutions are relatively small.
  The Amplitude from the first minimum at $t \simeq 1.7 \yr$ to the second maximum at $t \simeq 2.6 \yr$ of the regular resolution is 1.22 times larger than that of the high resolution. The average radius in the high-resolution is smaller. The ratio of this amplitude to the average radius at the minimum and second maximum in the regular resolution is 1.16 larger than that of the high resolution.   The oscillation period is larger in the regular resolution as the star expands more than in the high resolution; therefore, the dynamical time is larger.    
These small differences, $\simeq 20 \% $, indicate that it is sufficient to use the regular resolution when we add jets (as we have insufficient computer resources to simulate with the high resolution when we include jets).   

\begin{figure} 
\centering
\includegraphics[width=0.44\textwidth]{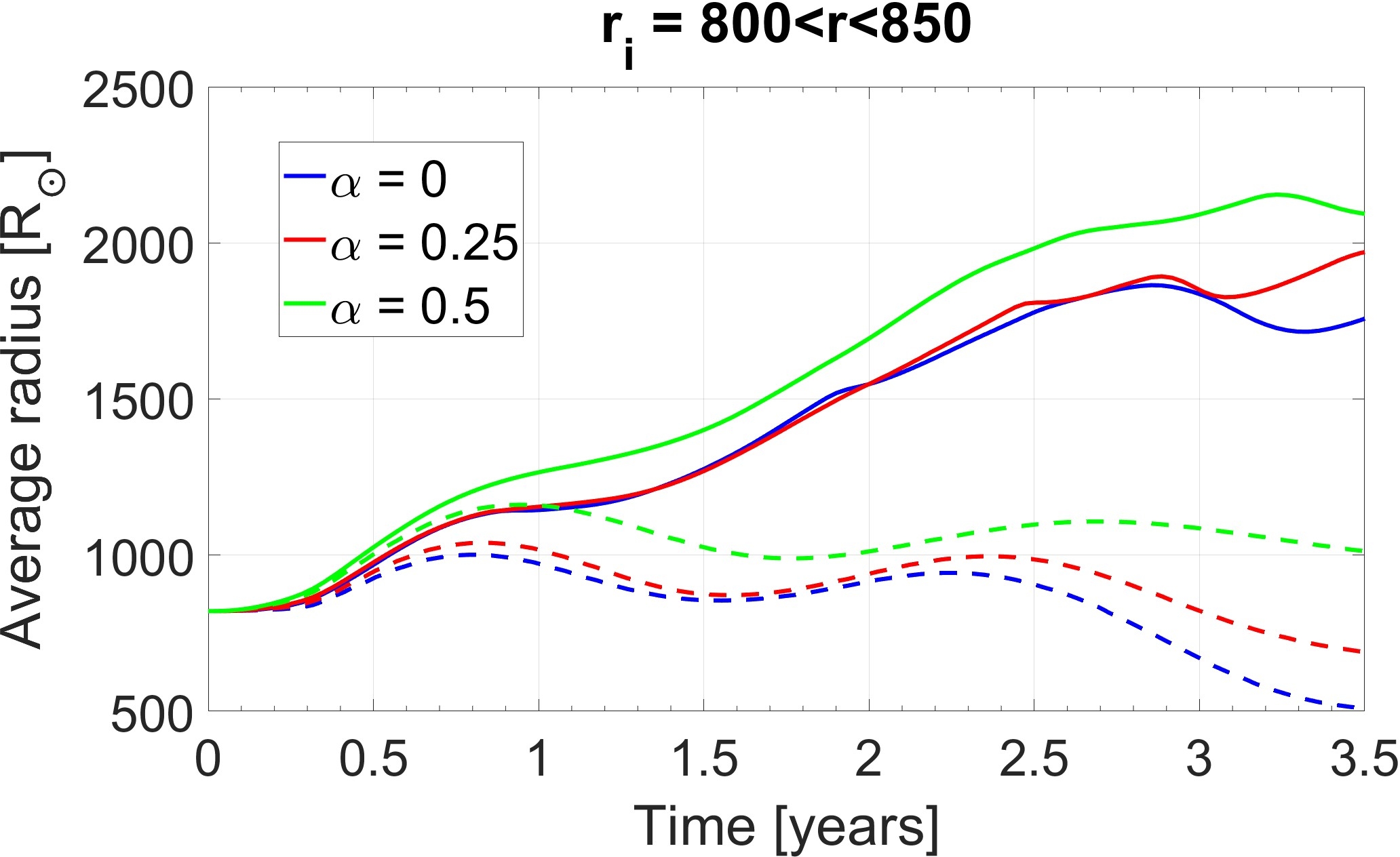}
\includegraphics[width=0.44\textwidth]{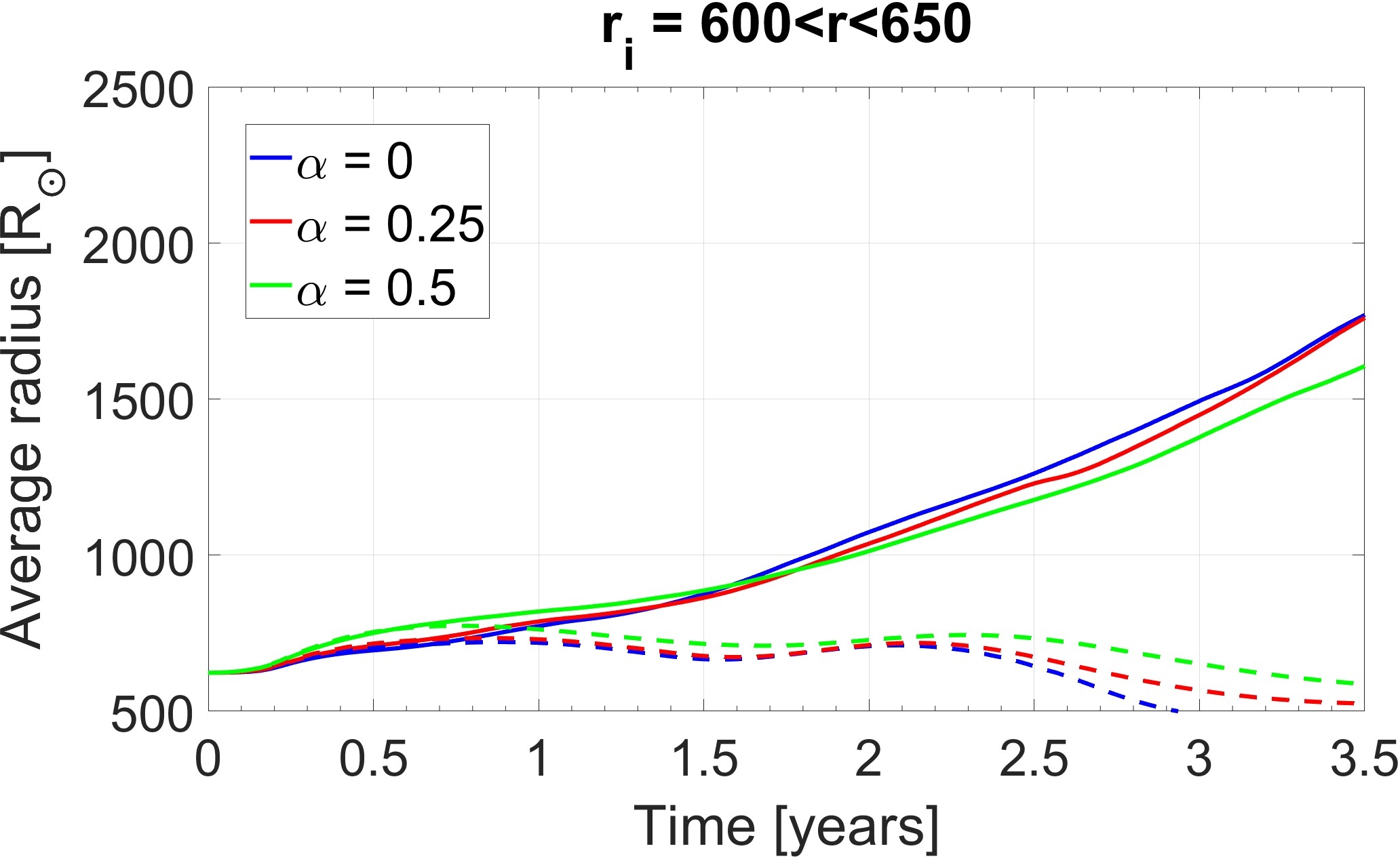}
\caption{The average radii of the tracers for two initial spherical shells: $800 R_\odot < r < 850 R_\odot$ (upper figure) and $600 R_\odot < r < 650 R_\odot$ (lower figure).
The solid lines represent simulations with jets, and the dashed lines represent those without jets. The insets show the initial envelope rotation according to equation (\ref{eq:Omega}). 
The Keplerian orbital period on the surface of the unperturbed RSG is $P_{\rm Kep}=2.35  \yr$, and the dynamical time is $P_{\rm D}\equiv (G \bar \rho)^{-1/2} =0.76  \yr$ where $\bar \rho$ is the average density of the initial RSG model. }
\label{fig:ShellsRadii}
\end{figure}

Figure \ref{fig:ShellsRadii} presents, by dashed lines in three colors, the average radius of the two shells for three simulations with different initial rotation velocities according to equation (\ref{eq:Omega}); all lines in Figure \ref{fig:ShellsRadii} are with regular resolution as they compare the simulations without jets with those with jets that we study in Section \ref{sec:Ejecta}. We learn that the rotating models also perform two oscillations before decaying. As expected, the centrifugal force increases the average radius. The larger radius increases the dynamical time and the oscillation period, as indicated by the peaks in the oscillating dashed lines.  We reiterate our claim that there is no need to relax the 3D stellar model; it can simply be allowed to oscillate. Introducing an energy source in the inner 3D stellar model that mimics the nuclear burning in the core might postpone the oscillation decay. This is the topic of ongoing research.  

  The solid lines in Figure \ref{fig:ShellsRadii} show that the jets do not prevent stellar oscillations. There are oscillations convolved with envelope expansion. The deposition of energy by the jets causes a significant expansion of the envelope, which increases the oscillation period. The oscillations are prominent in the outer tracer shell (upper panel of Figure \ref{fig:ShellsRadii}). The first maximum is at a time similar to that of the simulations without jets, but the radius is larger. The second minimum of the oscillation appears as a very shallow increase in radius. The second peak is clear in the $\alpha=0$ and $\alpha=0.25$ cases. Still, it occurs much later compared to the jetless simulations, as the envelope is large and the dynamic time is significantly longer.  

  Let us elaborate on the nature of the oscillations in the 3D stellar model and their relation to those of RSG stars. The driving energy of the oscillations in giant stars is the stellar luminosity resulting from nuclear burning in the core, such as the $\kappa$ mechanism, where opacity increases during compression and decreases during expansion. We do not yet have an energy source near the center of our 3D stellar model to mimic the stellar luminosity. Therefore, the oscillations we find here cannot be the same as those of RSG stars. 
The oscillations of the 3D stellar model result from the deviations of envelope zones from hydrostatic equilibrium. Indeed, these oscillations decay after about two periods. What we claim is that the oscillations resulting from the departure of the envelope from exact hydrostatic equilibrium mimic, in their periods and amplitudes, real RSG stellar oscillations more closely than a fully relaxed model does. We obtain prominent radial oscillations. The flow of the envelope mass in the non-rotating stellar models of \cite{Hilleletal2023} also exhibits tangential motions. The tangential motions are mainly the result of the large convective vortices as the envelope is also convective. Whether there are also non-radial pulsations unrelated to the convection is difficult to tell from our analysis and requires a separate study that better resolves the convective motion. Had we relaxed the model before starting the simulation, the stellar model would not oscillate, as evident from the decay of the oscillations. To have oscillations even in a relaxed model requires the numerical code to include an energy source near the center that mimics nuclear energy in the core, an envelope model with opacity that depends on the local temperature and density, and following ionization fraction.

\begin{figure*} 
\centering
\includegraphics[width=0.45\textwidth]
{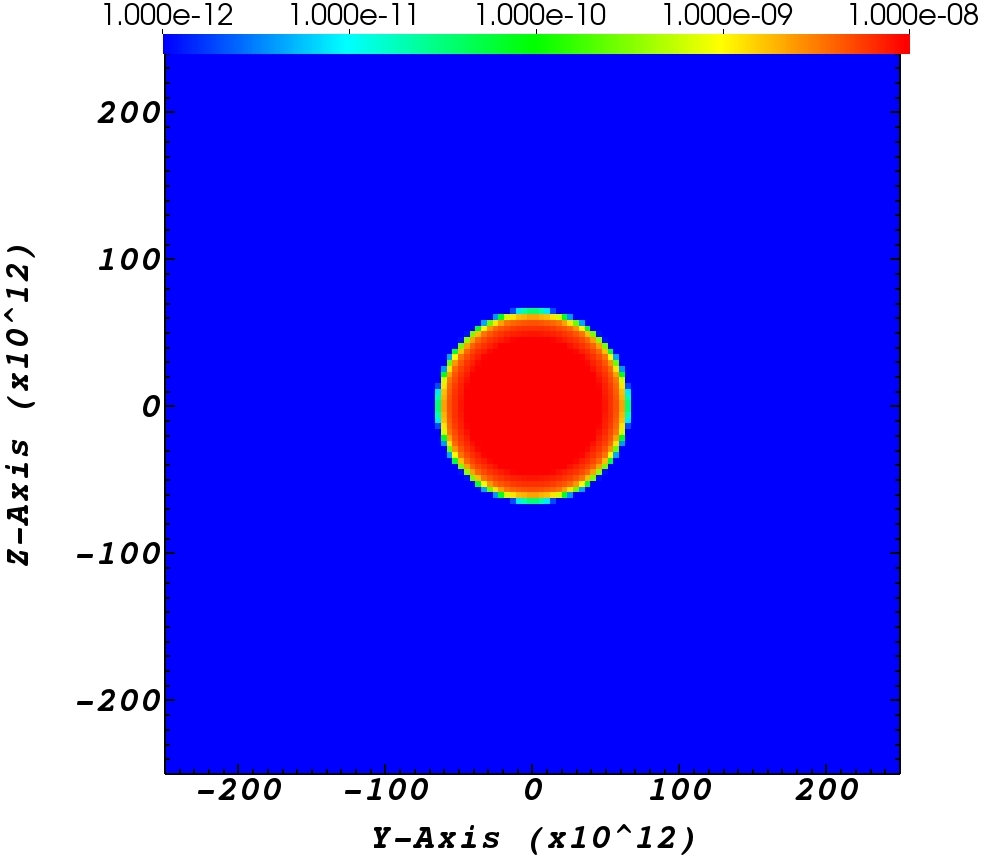}
\includegraphics[width=0.45\textwidth]
{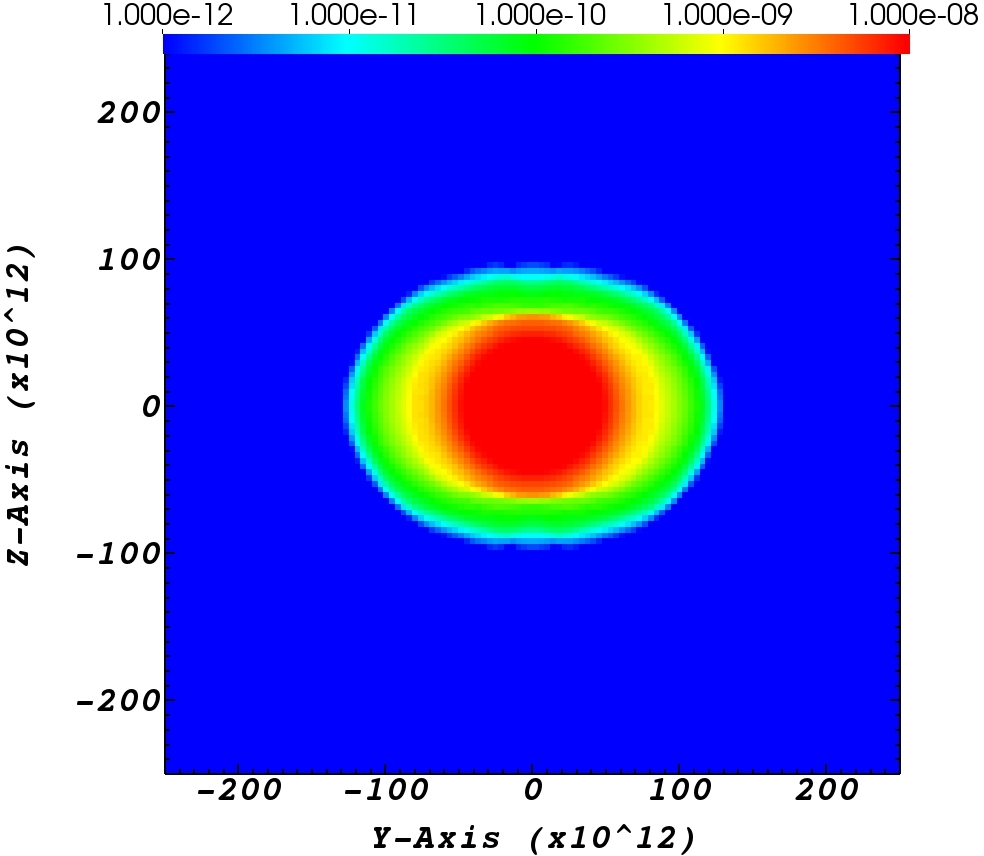}\\
\includegraphics[width=0.45\textwidth]
{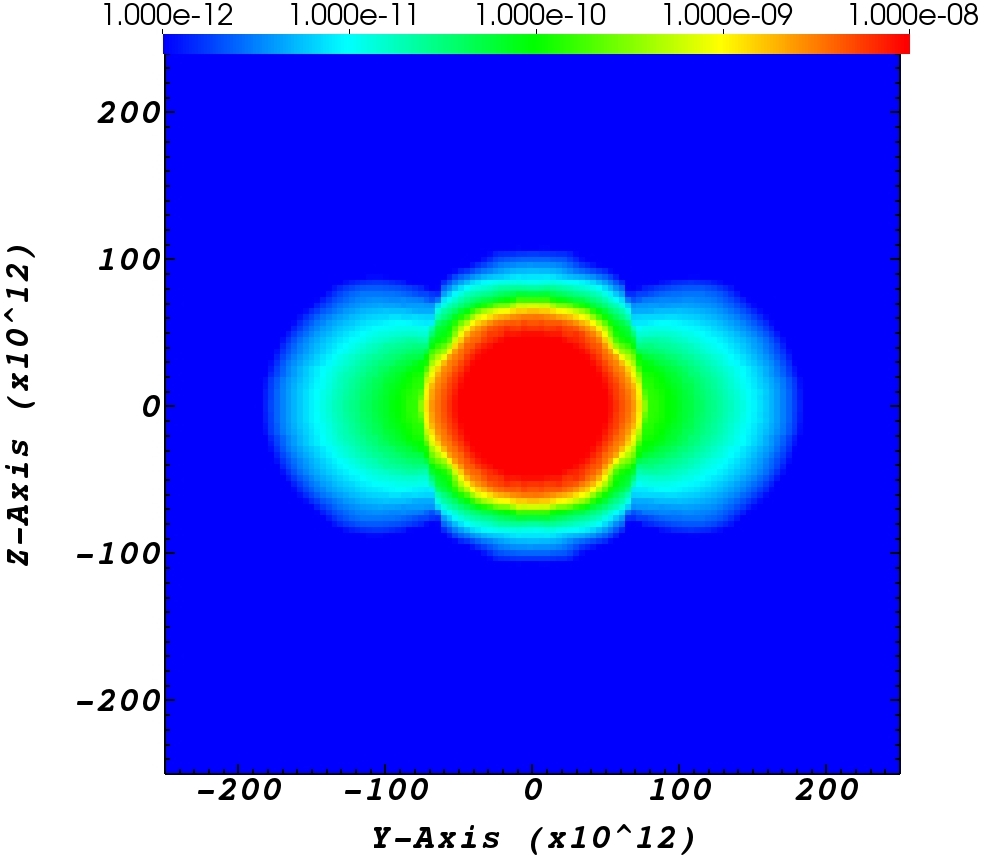}
\includegraphics[width=0.45\textwidth]
{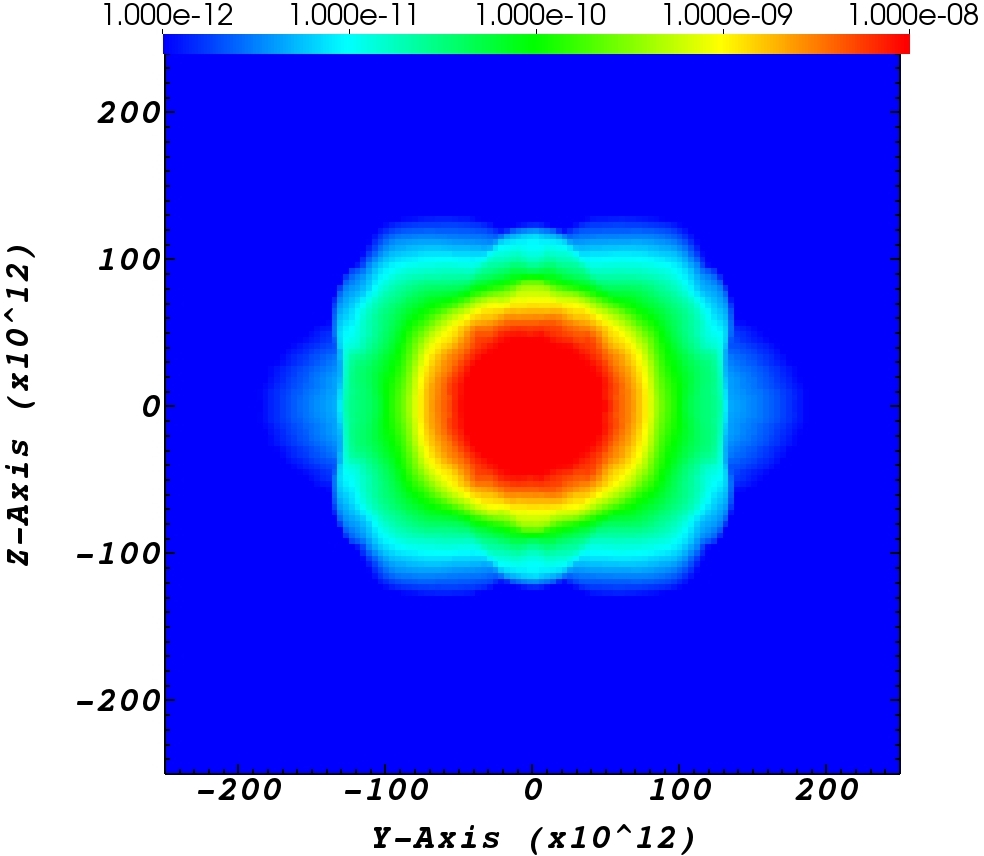}\\
\includegraphics[width=0.45\textwidth]
{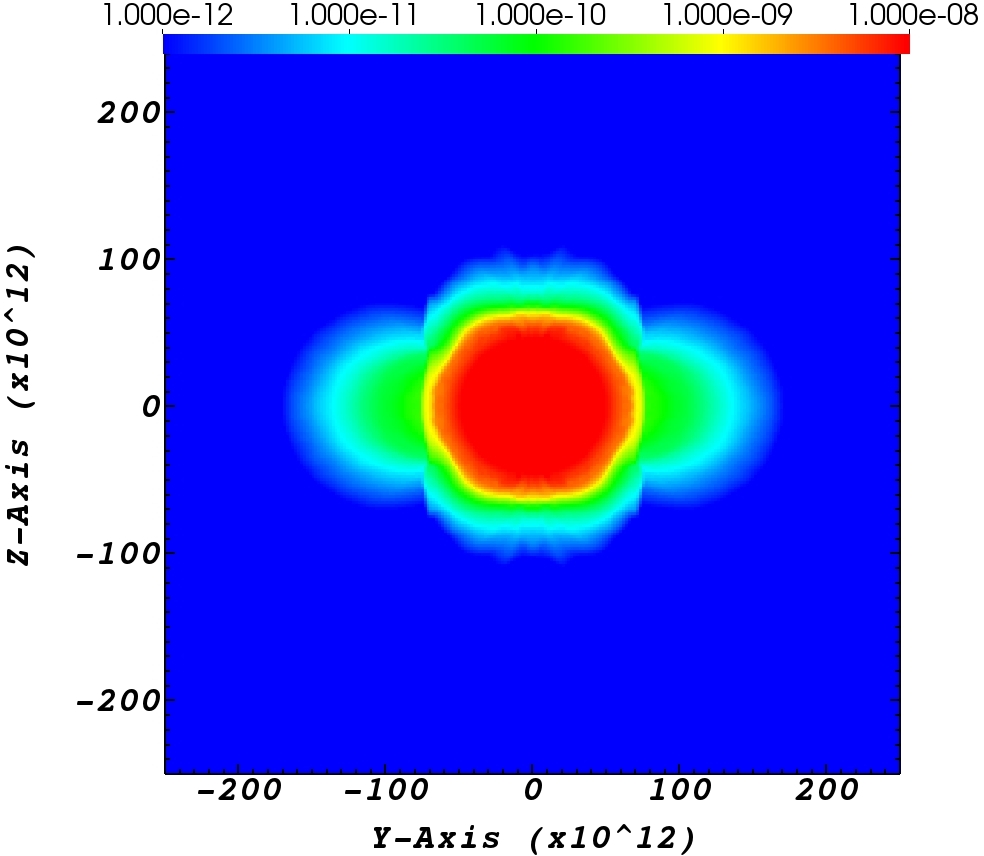}
\includegraphics[width=0.45\textwidth]
{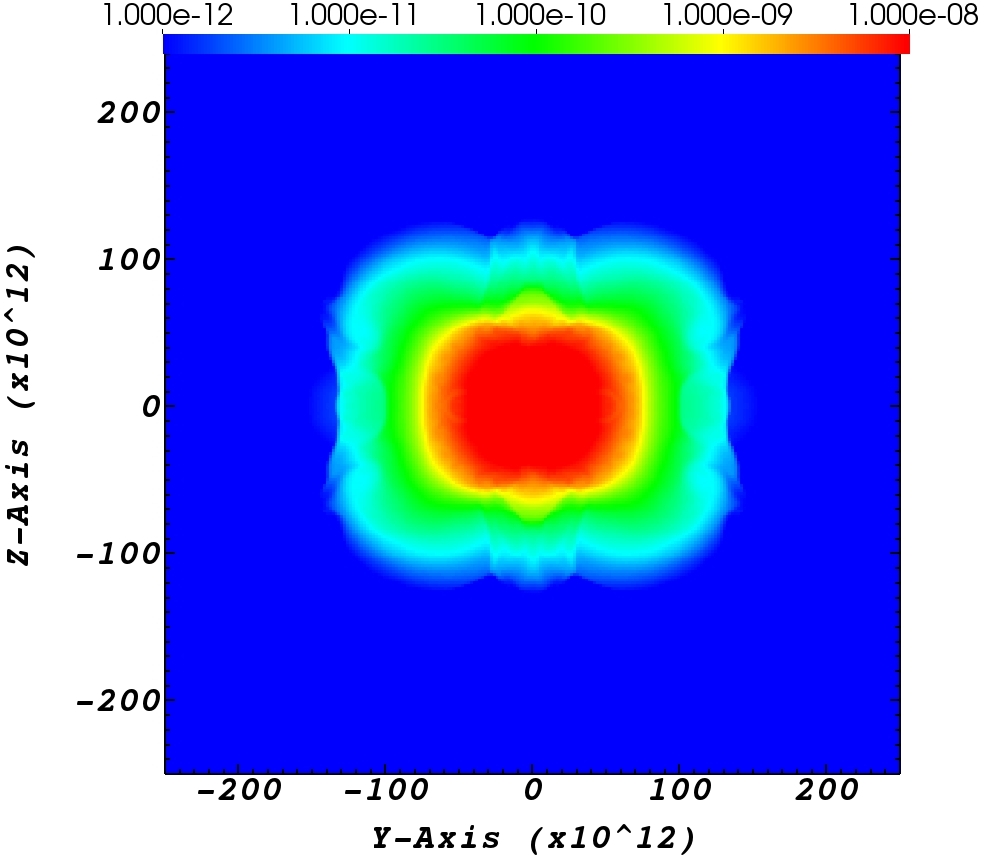}\\
\caption{Density maps in the $x=0$ meridional plane of the $\alpha=0.5$ no-jet simulations emphasizing the non-radial oscillations. The upper four panels are of the regular resolution at four times: $t=12\days$ (top left panel), at the first maximum of the oscillation according to the dashed-green line in Figure \ref{fig:ShellsRadii},  $t= 0.92 \yr$ (top right panel), at the first minimum of the oscillation, $t=1.8 \yr$ (middle left panel), and near the second maximum of the outer shell at $t=2.7 \yr$ (middle right panel). 
The lower panels are the high-resolution simulation results at $t=1.8 \yr$ (lower left) and at $t=2.7 \yr$ (lower right). 
Units of the axes are cm. The color bars depict the density from $10^{-12} \g \cm^{-3}$ (deep blue) to $10^{-8} \g \cm^{-3}$ (deep red). 
}
\label{fig:RotationDensity1}
\end{figure*}

Figure \ref{fig:RotationDensity1} presents the density map in the meridional plane $x=0$ at different times for the case $\alpha=0.5$. The four times of the first four panels of the regular resolution are practically at the beginning of the simulation, at first maximum, at the first minimum, and at the second maximum of the oscillation of the outer shell; this oscillation is depicted by the dashed-green line in the upper panel of Figure \ref{fig:ShellsRadii}. 
The two lower panels present the result of the high-resolution simulation of the same model at the same times as the middle row panels.  
As expected, the star expands more in the equatorial plane ($z=0$) than along the poles. Interestingly, the oscillation in the two orthogonal directions is out of phase. In moving from $t=1.8 \yr$ (lower left panels) to $t=2.7 \yr$ (lower right panels), on average, the star expands in the polar directions and contracts in the equatorial plane. The high and regular resolutions depict very similar behavior, justifying using the regular resolution when we include jets.  

\begin{figure*} 
\centering
\includegraphics[width=0.33\textwidth]{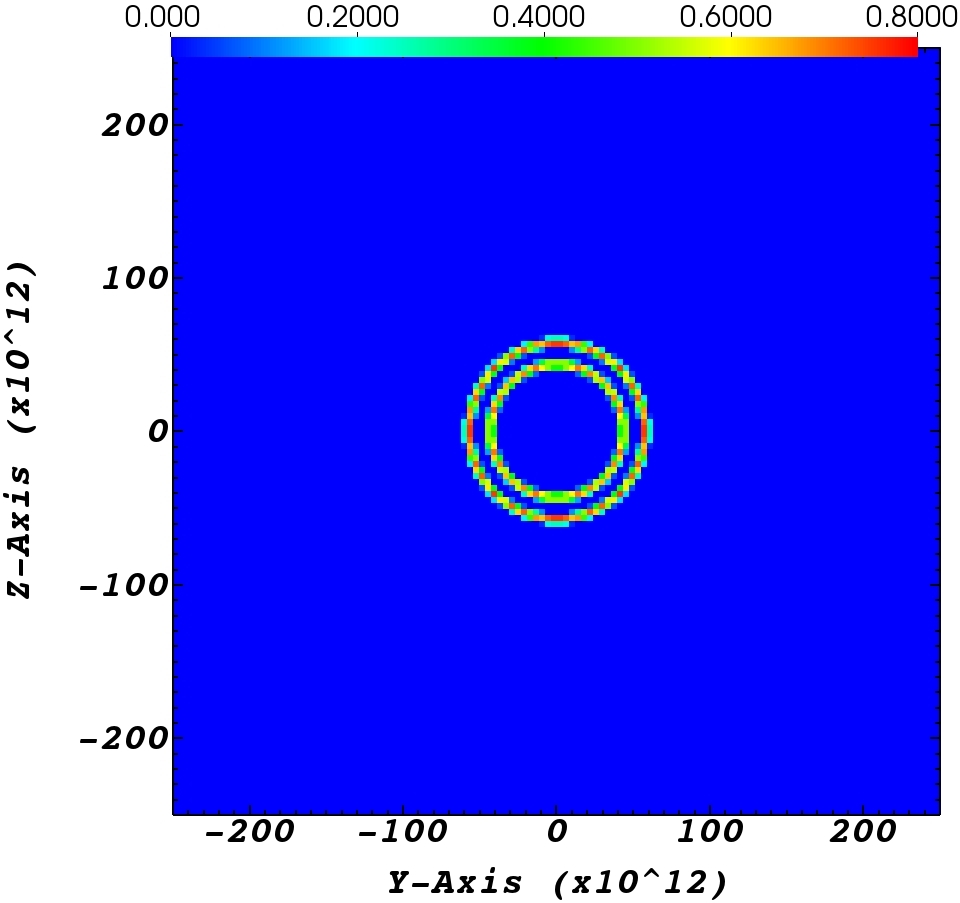}
\includegraphics[width=0.33\textwidth]{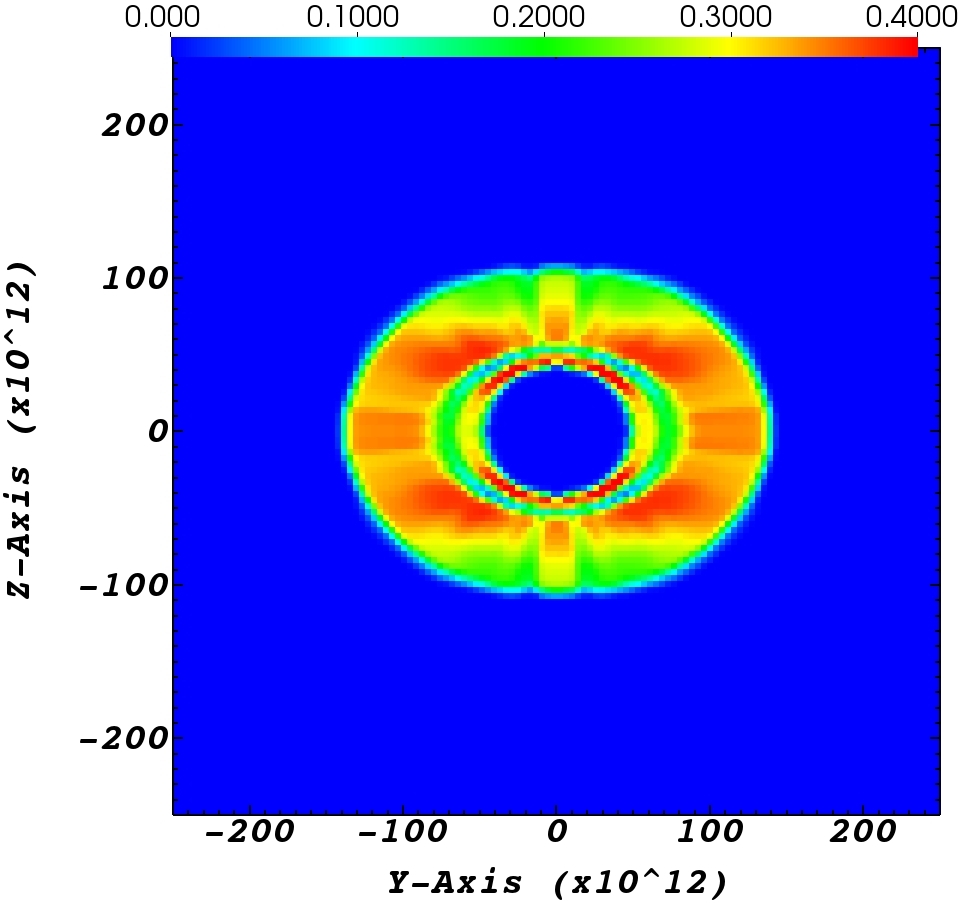}\\
\includegraphics[width=0.33\textwidth]{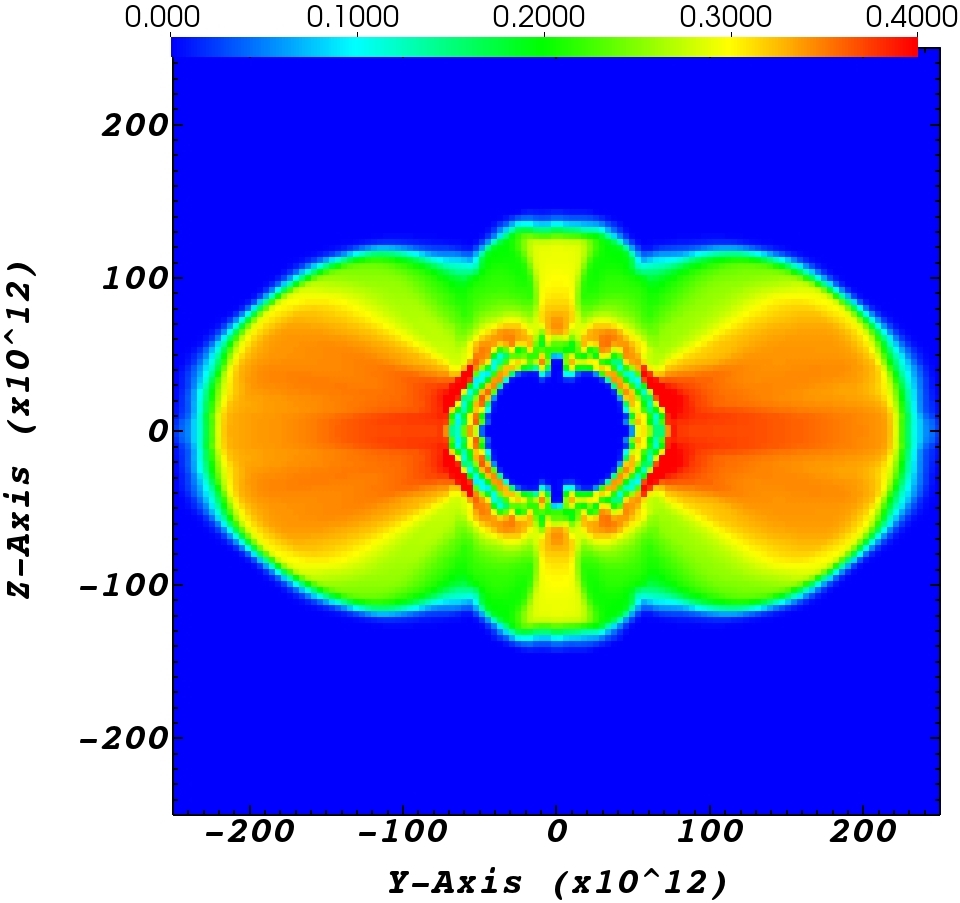}
\includegraphics[width=0.33\textwidth]{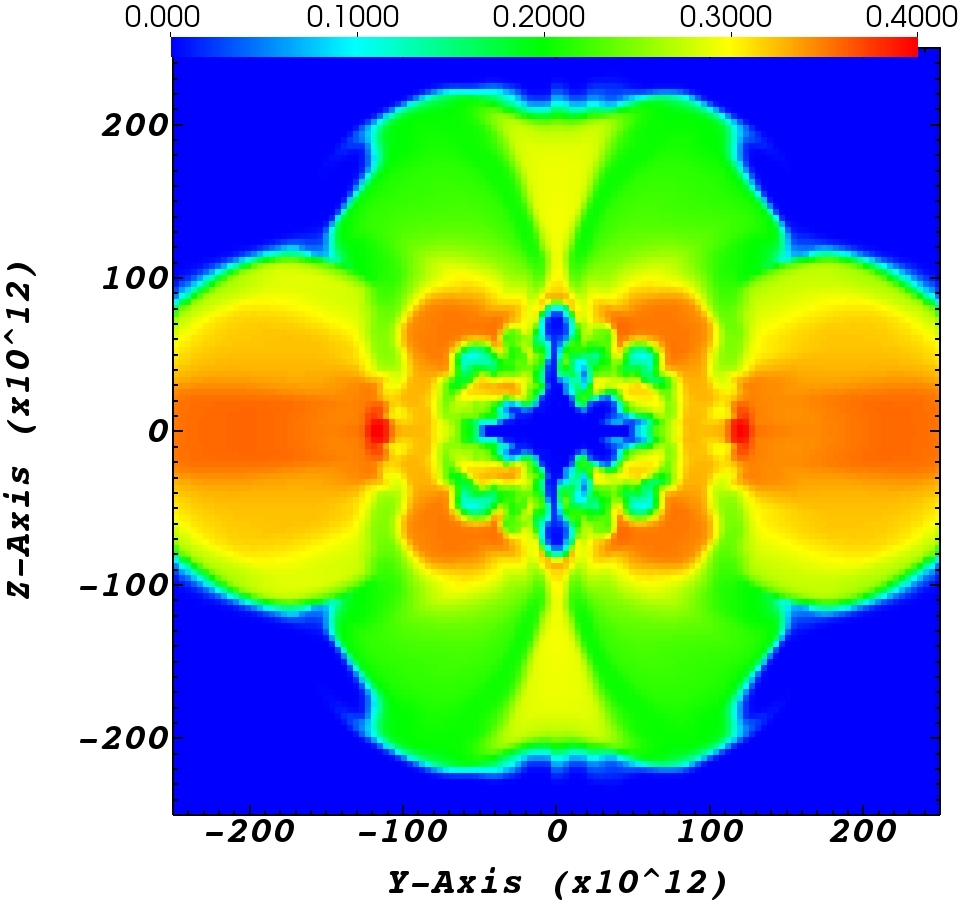}\\
\includegraphics[width=0.33\textwidth]{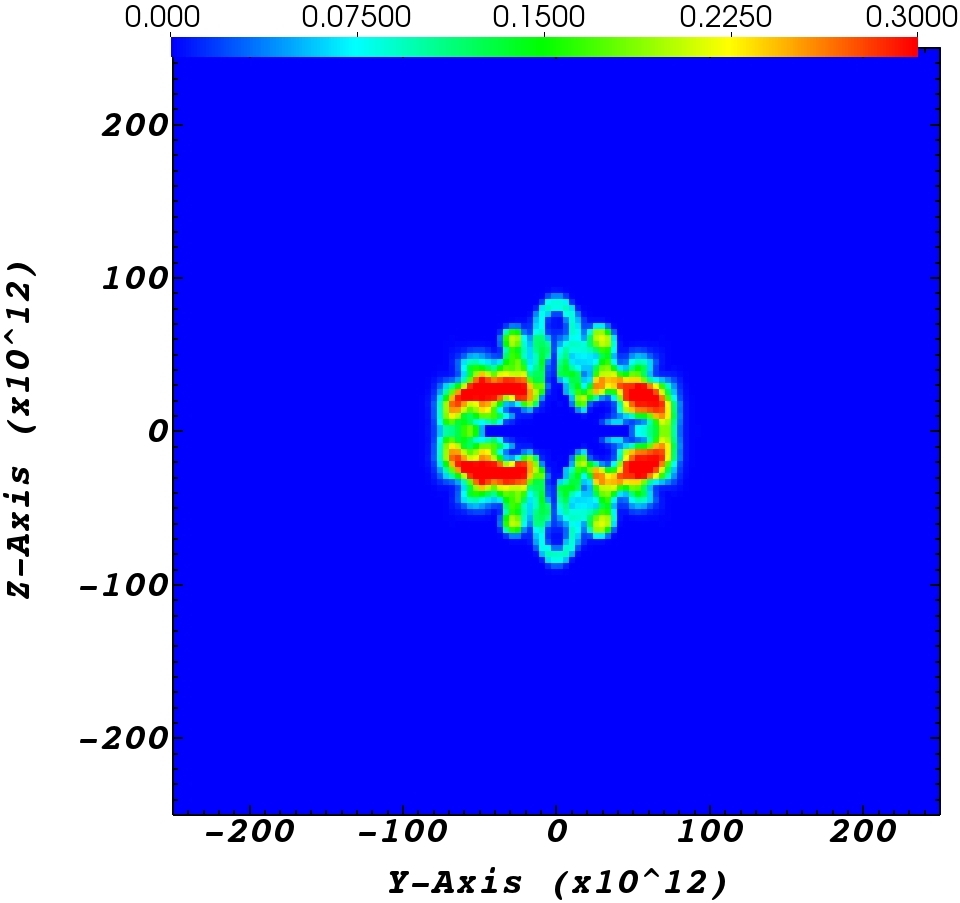}
\includegraphics[width=0.33\textwidth]{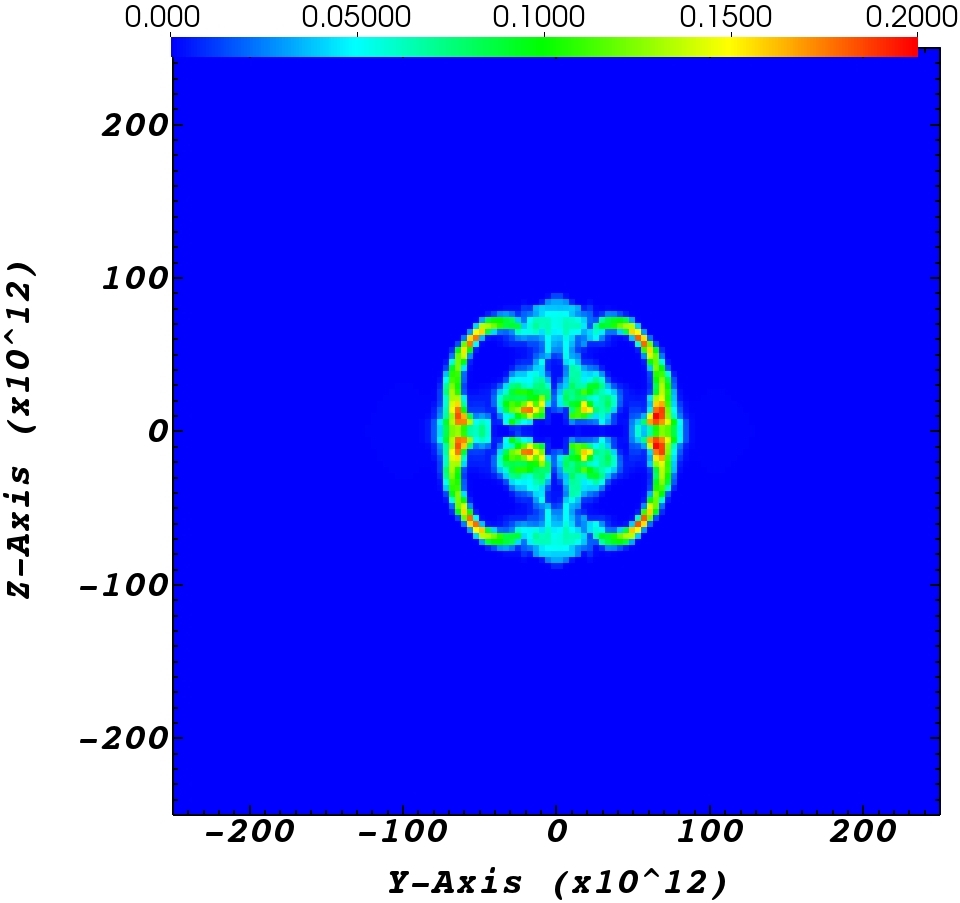}\\
\includegraphics[width=0.33\textwidth]{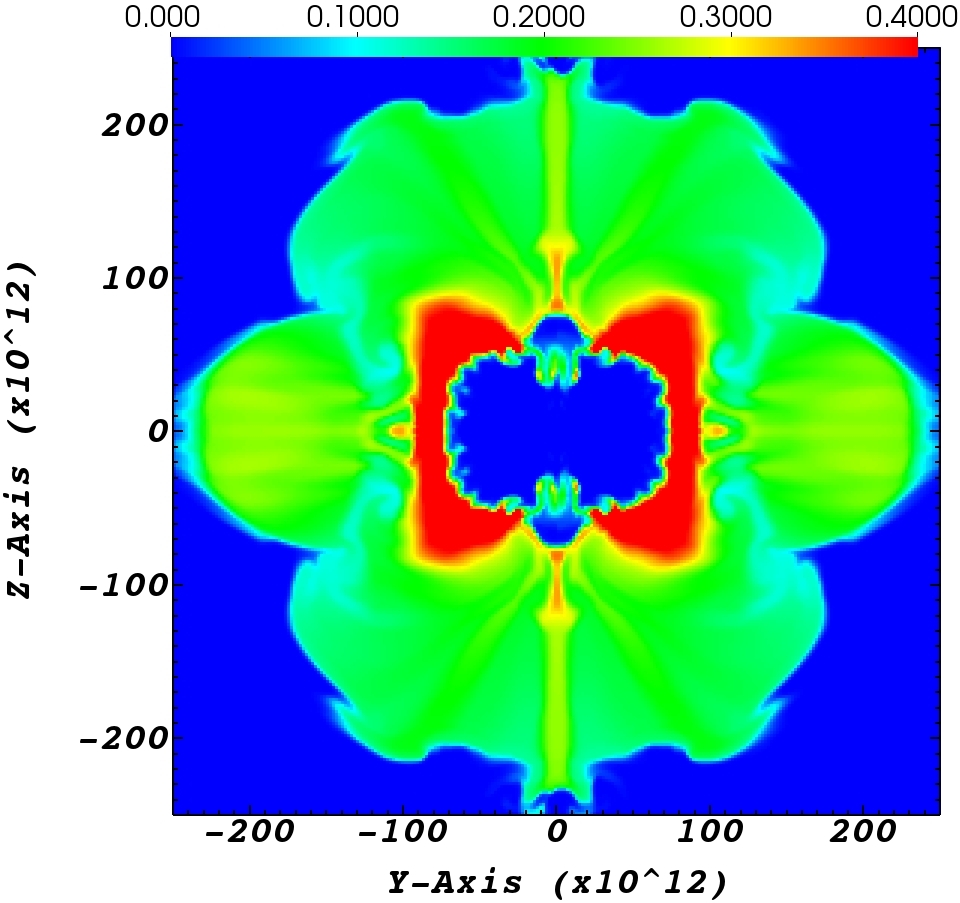}
\includegraphics[width=0.33\textwidth]{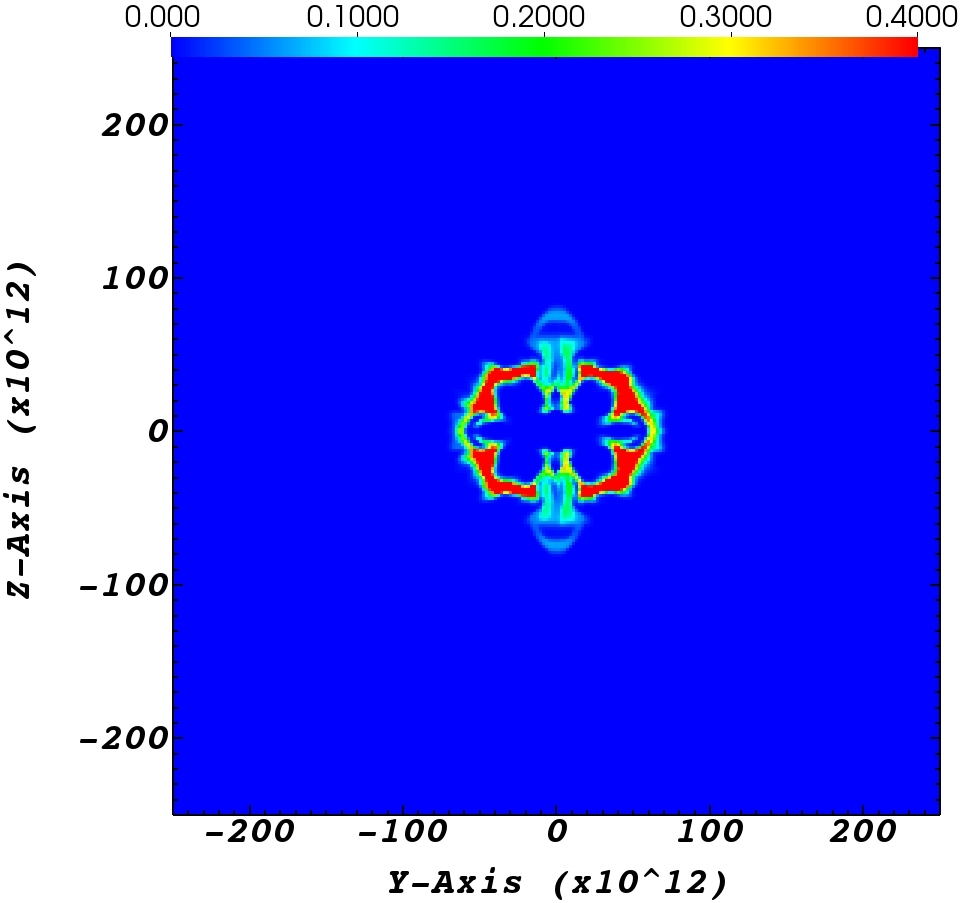}
\caption{Tracer maps of two shells in the meridional $x=0$ plane
at $t=12\days$ (top left), $t=0.92 \yr$ (first maximum in the oscillator motion; top right panel), $t=1.8 \yr$ (first minimum; second left),  and $t=2.7 \yr$ (second right).
The two panels in the third row present only the tracer of the inner shell at
$t=2.7 \yr$ (left), and $t=3.5 \yr$ (right), at about the dynamical time. 
The upper six panels are of the regular simulation with $\alpha=0.5$. 
The last two panels show the tracers of the outer (left) and inner (right) shells at $t=2.7 \yr$ of the high-resolution simulation with $\alpha=0.5$. 
The units on the axes are cm. The tracer scales are on top of the panels and change between panels. }
\label{fig:RotationTracers}
\end{figure*}

The upper four panels of Figure \ref{fig:RotationTracers} present the tracer maps in the meridional plane $x=0$ of the two shells that we show in Figure \ref{fig:ShellsRadii} for the $\alpha = 0.5$ simulation, and at four times as in Figure \ref{fig:RotationDensity1}, including minima and maxima of the average outer shell radius. The upper left panel presents the two shells at the beginning of the simulation ($t=12\days$). The two shells are well separated with envelope gas free of the shells' material. At $t=0.92 \yr$, the first maximum in the oscillation, the envelope flattens, as expected from the centrifugal force, and the shells' material mixes with the region between the shells. This mixing is expected as the envelope is convective, as demonstrated in the non-rotating case in the previous paper \citep{Hilleletal2023}.  As we simulate only the middle and outer zones of the RSG envelope, and not the inner inert sphere, at $t=0$, the entire envelope (beside a very thin layer near the surface) is convective.   While the inner shell maintains its identity, the outer shell smears to larger radii than its initial radius. The left panel in the second row at $t=1.8 \yr$ shows that at the minimum of the oscillation, some material from the outer shell spreads to the edge of the grid while some outer shell material rebuilds the shell near the equatorial plane. The inner shell continues to mix with its surroundings (color change from deep red at $t=0.92 \yr$ to yellow or faint red) but still maintains its identity.  The right panel in the second row shows the continuation of this evolution.  
The two panels in the third row of Figure \ref{fig:RotationTracers} show the inner shell at $t=2.8 \yr$ (second maximum) and at $t=3.5 \yr$, respectively.  We can see that throughout the entire simulation, the inner shell remains confined. 
The lower two panels present tracer maps of the high-resolution simulation at $t=2.7 \yr$: lower left: the tracer of the outer shell; lower right: the tracer of the inner shell.   
The shells' behavior shows that although the outer part expands and its material flows out of the grid boundary, the inner parts stay inside the star. This behavior implies that we can trust the simulations if the companion spirals into the envelope. First, the energy the jets deposit significantly affects mass ejection and envelope disruption more than oscillation and numerical inaccuracies. Second, in a short time, the NS spirals to the center, where the model stays intact for several years, which is several dynamical times, more than we need for our present study with jets (Section \ref{sec:Ejecta}).    

\begin{figure*} 
\includegraphics[width=0.49\textwidth]{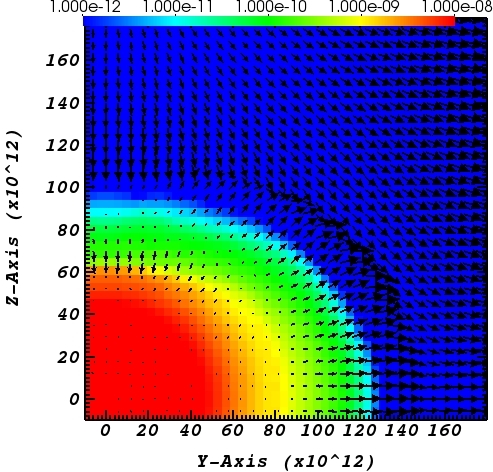} 
\includegraphics[width=0.49\textwidth]{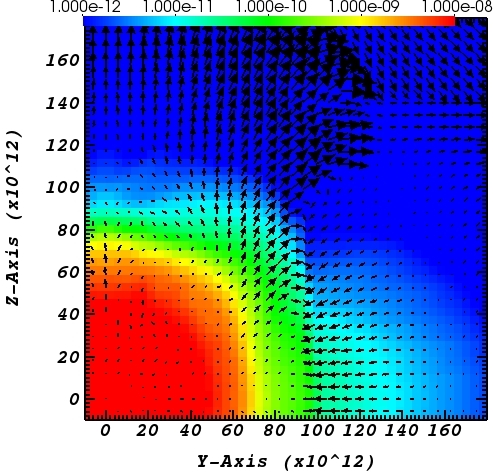}\\
\includegraphics[width=0.49\textwidth]{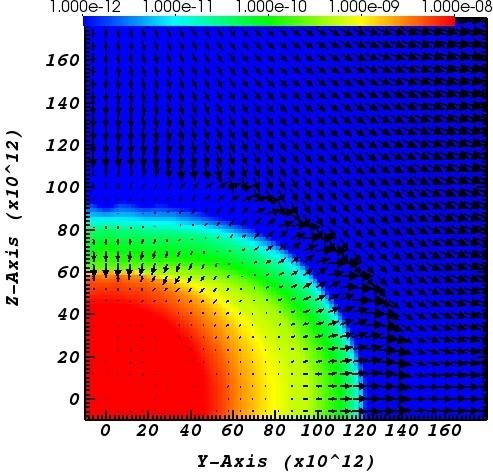} \includegraphics[width=0.49\textwidth]{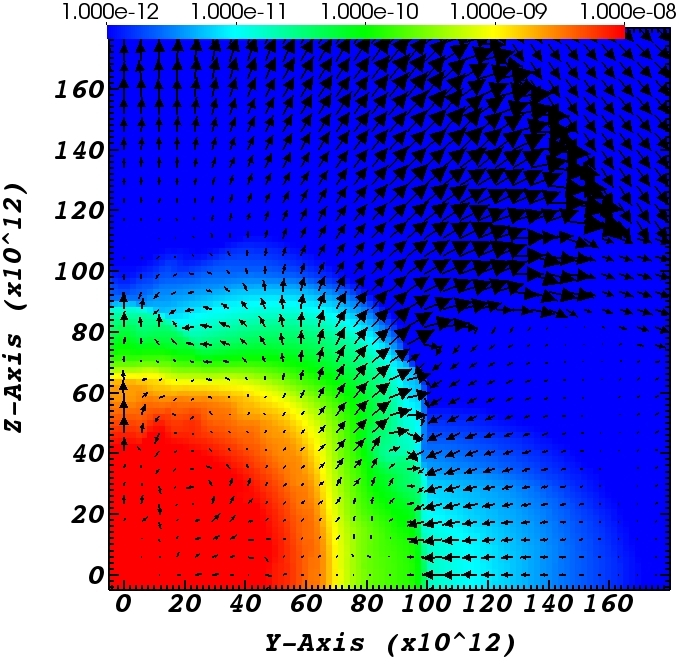}
\caption{Velocity vectors on top of density maps in a quarter of the meridional $x=0$ plane at $t=0.92 \yr$ (left) and at $t=2.3 \yr$ (right). Note that we simulate the entire volume, but for clarity, present only one-quarter of the plane.   
The upper and lower panels are for the regular and high-resolution simulations for $\alpha=0.5$. 
Units on the axes are cm, and the largest arrow
at $t=0.92 \yr$ and at $t=2.3 \yr$
corresponds to velocities of $90 \km \s^{-1}$
and $50 \km \s^{-1}$, respectively. The color bars depict the density from $10^{-12} \g \cm^{-3}$ (deep blue) to $10^{-8} \g \cm^{-3}$ (deep red).}
\label{fig:Dens_Vel_vectrQ}
\end{figure*}

Figure \ref{fig:Dens_Vel_vectrQ} presents the velocity maps in a quarter of the meridional plane $x=0$ and at two times, for the regular (upper panels) and high (lower panel) resolution no-jet simulations. The blue zone is of no interest to us as it is the low-density medium that we insert around the star at $t=0$ (and the arrows are hard to see). This figure emphasizes the non-radial oscillation. At $t=0.92 \yr$ (left panels), the material along the polar direction near $z=60 \times 10^{12} \cm$ moves inward (contracts), while material near the equatorial plane ($y \simeq 110 \times 10^{12} \cm$ in the maps) expands. At $t=2.3 \yr$ (right panels), the velocities in these regions are in the opposite directions. There are large vortices that are clearer in the high-resolution simulation. These four maps also show that the differences between the regular and high-resolution simulations are not large.  

\begin{figure} 
\centering
\includegraphics[width=0.46\textwidth]{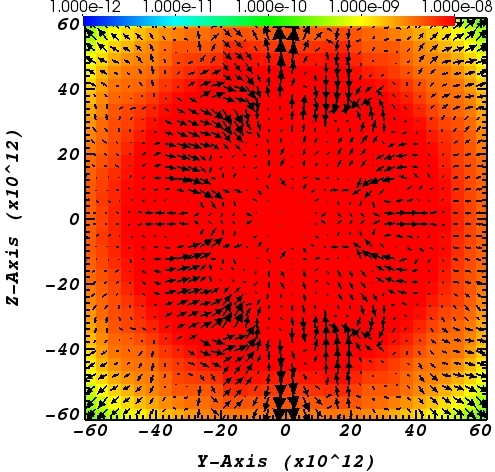}\\
\includegraphics[width=0.46\textwidth]{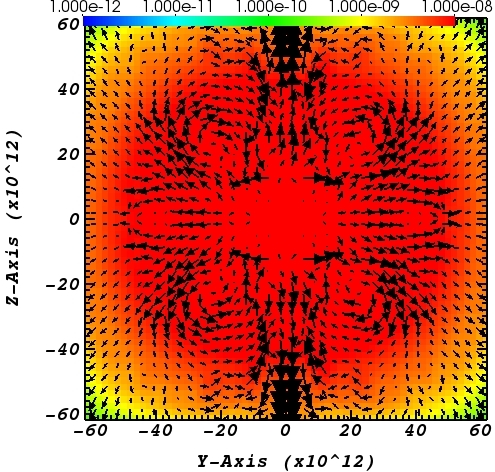}
\caption{Velocity vectors on top of density maps in the meridional $x=0$ plane at $t=2.3 \yr$ 
for the regular-resolution simulation (top) and the high-resolution simulation (bottom) with $\alpha=0.5$.
The largest arrow corresponds to a velocity of $53 \km \s^{-1}$.
}
\label{fig:Dens_Vel_vectrFull}
\end{figure}

Figure \ref{fig:Dens_Vel_vectrFull} presents the velocity map in the meridional $x=0$ plane extending to the initial radius of the star for the regular (upper panel) and high (lower panel) resolution simulations. The velocity maps present vortices (clearer in the high-resolution map in the lower panel), as expected in a convective envelope, and as we found for the non-rotating model \citep{Hilleletal2023}. The figure presents another interesting pattern of the vortices. There is an almost perfect mirror symmetry about the equatorial plane $z=0$. However, the two sides of the meridional plane, i.e., $y<0$ and $0<y$ are not symmetric. Probably, some kind of instability due to the rotation amplifies small perturbations that the numerical code introduces. Namely, the perturbation is along the tangential direction because of the rotation.  

Although the high resolution is more accurate, we concentrated on the simulations with the regular resolution because when we add jets, the time steps are substantially reduced, and we do not have the computer power to follow jetted simulations with the high-resolution grid. Our comparison of the no-jet simulations shows that the regular resolution is adequate for our goals even when we set envelope rotation.  

With the preset rotating stellar model, we find the convection and oscillation of two periods before decay, as we found and discussed in detail in the non-rotating model \citep{Hilleletal2023}. The new finding here is that the oscillations are non-radial. These non-radial oscillations of the rotating model deserve a separate study.  

We can summarize the simulations of the stellar model before the onset of the CEE by strengthening our claim from  \cite{Hilleletal2023} that \textit{there is no need to stabilize and relax the stellar model before starting the CEE simulation.} The very large oscillations and the mixing in the envelope that results from the turbulent convection are actual processes. RSG stars have strong envelope convection and prominent amplitude pulsations. The radius variation (maximum radius minus minimum radius in an oscillation cycle) of observed long-period variable stars can be of the order of the stellar radius $\Delta R \simeq R$ (e.g., \citealt{Trabucchietal2021}).
Our code is inaccurate in calculating the very outer parts of the stellar model after about a dynamical time $P_{\rm D}\equiv (G \bar \rho)^{-1/2} =0.76  \yr$.   The code does much better and for several dynamical times, which is about five years, in most of the envelope. Since we launch very energetic jets as we start the simulations, with no delay time, the effect of the jets in expelling the outer parts of the envelope is much larger than the numerical inaccuracy in the outer parts. In other words, the jets expel the outer parts of the envelope before the numerical inaccuracies develop.    
 
\section{Bubbles and filaments in the ejecta} 
\label{sec:Ejecta}

This section describes the evolution when we launch jets from the position of the NS. The simulations do not explicitly include the gravity of the NS and reveal the sole roles of the jets.  Namely, the main role of the NS gravity is in accreting mass and launching jets. The NS gravitational force on the envelope is not taken into account. We explore the role of jets in this CEE process.     
We pre-set the orbit of the NS to spiral-in from $a_{\rm i} = 850\,R_{\odot}$  to $a_{\rm SR} = 300\,R_{\odot}$ in a timespan of 3 years. It then performs a circular orbit at radius $a_{\rm SR}$.   The reason we stop the spiraling-in of the NS at $a_{\rm SR} = 300\,R_{\odot}$ is numerical. To conserve computational resources, we do not simulate the evolution within the inner $20\%$ of the envelope, specifically in the inert core, which has a radius of $R_{\rm inert}=176 R_\odot$. We cannot deposit the energy of the jets too close to this radius and, therefore, do not allow the NS to spiral in below $a_{\rm SR} = 300 R_{\odot}$.   The radial velocity during the rapid in-spiral (plunge-in) phase is constant, while the tangential velocity is the Keplerian one at each radius (details are in \citealt{Hilleletal2023}). The power of the jets is a small fraction of the power that a Bondi-Hoyle-Lyttleton accretion releases; the small factor results from the negative jet feedback mechanism; the power of the jets increases from $\dot E_{2j} (a_{\rm i}) = 1.6\times 10^{41} \erg \s^{-1}$ at the outer orbital separation to $\dot E_{2j} (a_{\rm SR}) = 1.4\times 10^{42} \erg \s^{-1}$ at the final orbit (details are in \citealt{Hilleletal2023}).

\begin{figure*} 
\centering
\includegraphics[width=1.0\textwidth,trim={0 7cm 0 0},clip]{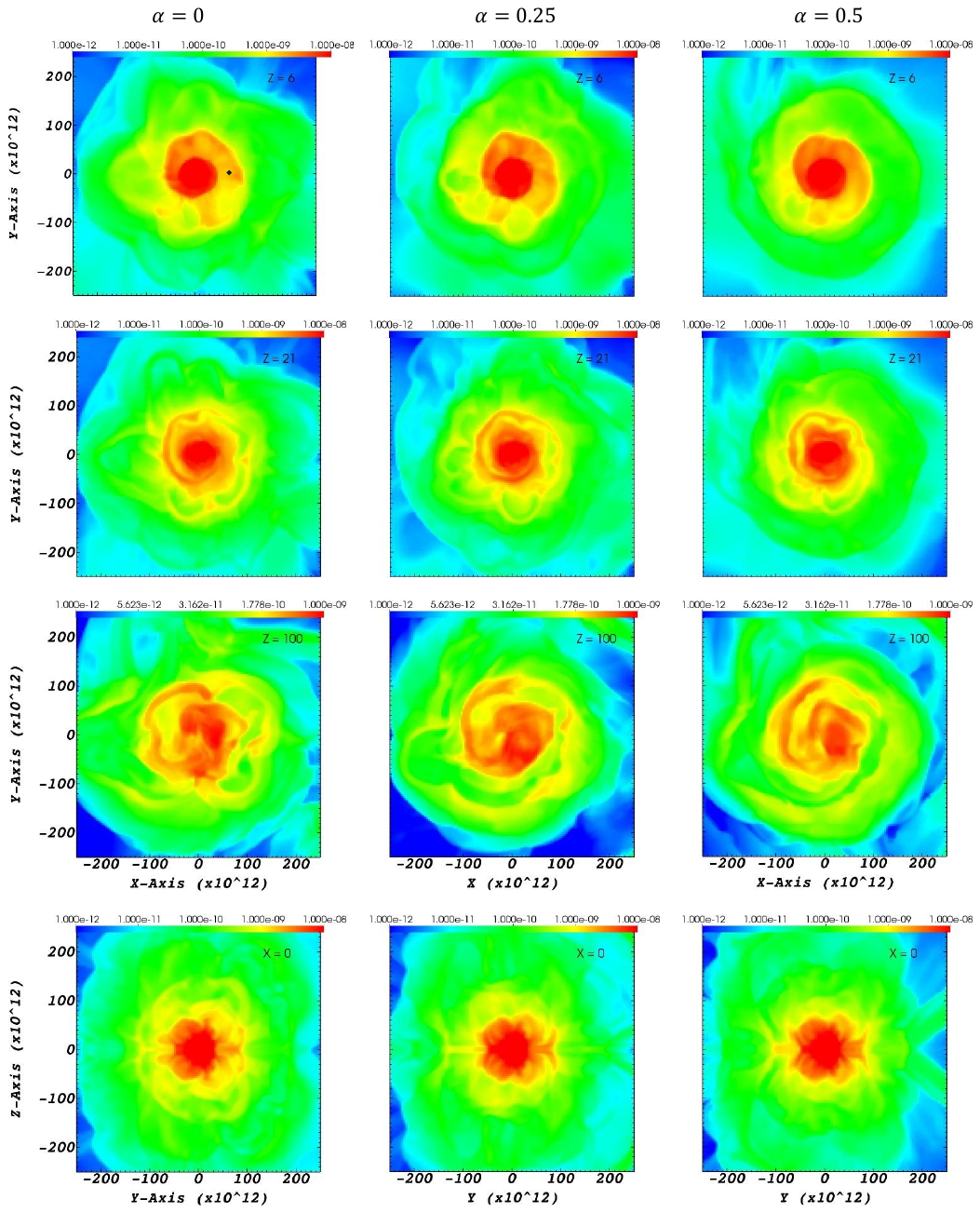}
\caption{Density maps of the three simulations 
with $\alpha=0$, $0.25$ and $0.5$, left to right, at $t= 3.8 \yr$. 
The first three rows are for different planes parallel to the equatorial plane; the insets show the distance of each plane from the equatorial plane in units of $10^{12} \cm$, namely, $z=6\times 10^{12} \cm$, $21\times 10^{12} \cm$, and $100\times 10^{12} \cm$. The lower row shows the density maps in the plane $x=0$. 
The upper left panel shows a black dot marking the initial (at $t=0$) NS location.
These density maps emphasize stronger spiral structures for faster rotation and the formation of medium-size structures, i.e., filaments and bubbles. 
}
\label{fig:DensityMaps}
\end{figure*}

Figure \ref{fig:DensityMaps} compares the density maps in two planes at three times of three simulations with jets: the left column is the non-rotating simulation as in \cite{Hilleletal2023}, the middle and right columns are new simulations for a rotating envelope at $t=0$ with values of $\alpha=0.25$ and $\alpha=0.5$, respectively, where we defined $\alpha$ in equation (\ref{eq:Omega}). The spiral structure in the planes parallel to the equatorial plane (upper three rows) becomes more prominent with increasing initial envelope rotation  (moving from left to right). Other than this, the general qualitative appearance of the density maps is similar in the three simulations. Therefore, we will not repeat the study of the ejecta morphology, the shedding of pairs of vortices in an expanding spiral pattern as the NS spirals-in, and the entropy profiles that we explored for the non-rotating simulation in \cite{Hilleletal2023}. 

Figure \ref{fig:ShellsRadii} shows that the large energy that the jets deposit into the envelope causes its expansion to the degree that the shells we follow do not perform oscillations (solid lines in the two panels of the figure).

\begin{figure*} 
\centering
\includegraphics[width=1.0\textwidth,trim={0 7cm 0 0},clip]{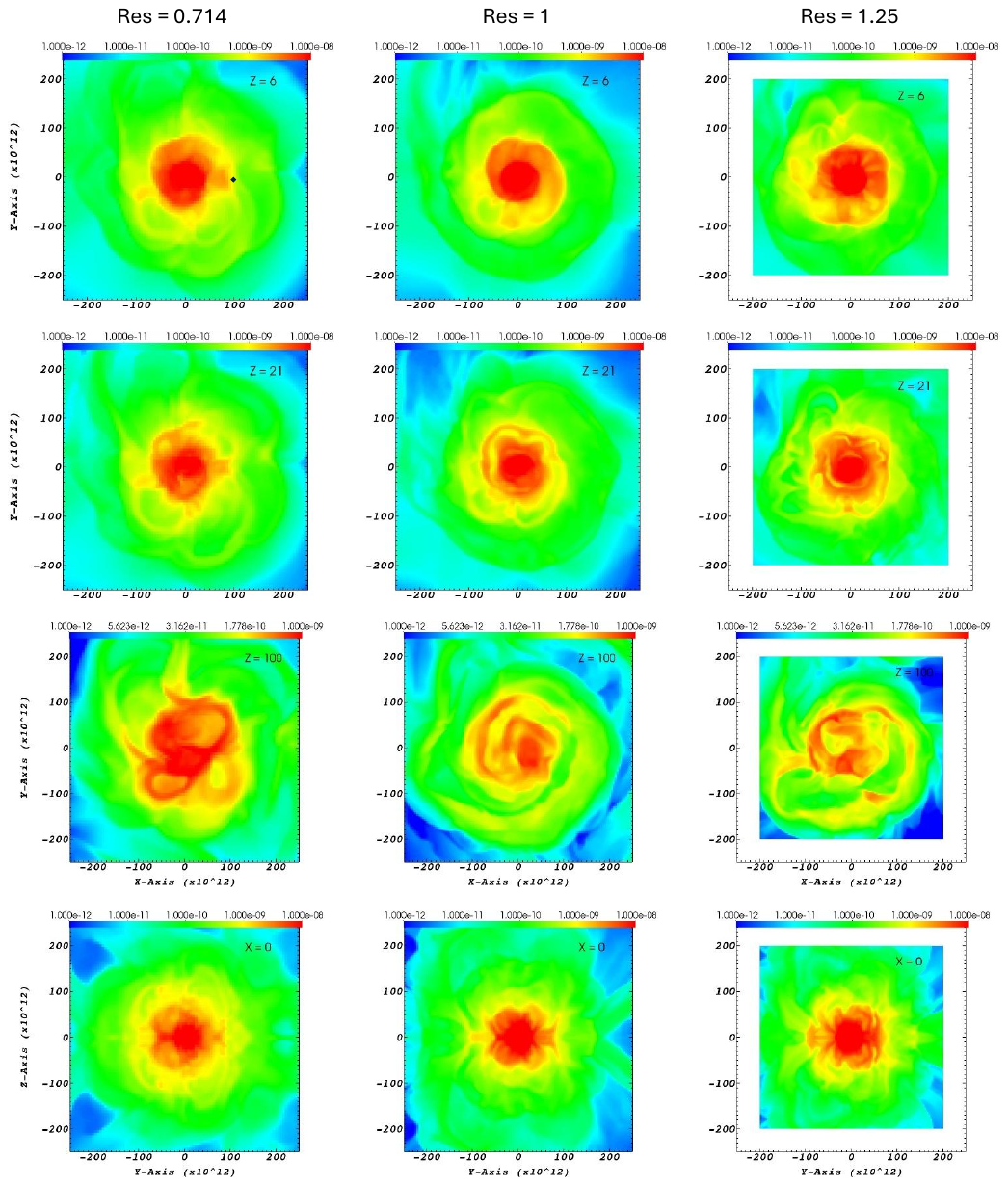}
\caption{  Density maps of three simulations with different resolutions, all with $\alpha=0.5$. The density maps are at $t= 3.8 \yr$ and in the same four planes as in Figure \ref{fig:DensityMaps}.  
The first three rows are for different planes parallel to the equatorial plane; the insets show the distance of each plane from the equatorial plane in units of $10^{12} \cm$, namely, $z=6\times 10^{12} \cm$, $21\times 10^{12} \cm$, and $100\times 10^{12} \cm$. The lower row shows the density maps in the plane $x=0$. 
The upper left panel shows a black dot marking the initial (at $t=0$) NS location.
The middle column is the same as the right column of Figure \ref{fig:DensityMaps}, where all cells have a size of $\Delta_{\rm cell}=3.90625 \times 10^{12} \cm$. The left column has a lower resolution with a larger grid-cell size by a factor of $1/0.714=1.4$. The right column has a higher resolution with a grid-cell size smaller by a factor of $1/1.25=0.8$; the grid is smaller due to the small cell size.     
}
\label{fig:DensityMapsResolution}
\end{figure*}

In Figure \ref{fig:DensityMapsResolution} we compare the density maps of three simulations with three resolutions for $\alpha=0.5$, and at the same time and planes as in Figure \ref{fig:DensityMaps}. The middle column of Figure \ref{fig:DensityMapsResolution} is the same as the right column of
Figure \ref{fig:DensityMaps}; the left column is for lower resolution and the right column is for higher resolution. The lower resolution (left column) does not show a clear spiral pattern. Other than this, all simulations show the formation of filamentary extended envelope and ejecta. 
 
A new analysis of the present study is the role of the Rayleigh-Taylor Instability.  
The most prominent pattern in the density maps (Figure \ref{fig:DensityMaps}) common to the three simulations is the appearance of dense filaments and low-density bubbles. As we show below, Rayleigh-Taylor instability modes influence the development of these medium-size structures, i.e., the typical sizes of the filaments and bubbles are below the radius of the star but only by a factor of a few. 

The enormous energy that the jets deposit in the vicinity of the NS, relative to the thermal energy of the envelope gas in this region, forms a high-pressure zone that expands, lowering its density. The flow structure where a low-density high-pressure zone accelerates envelope regions around it is prone to Rayleigh-Taylor instability modes. When the angle between the density and pressure gradient is larger than $90^\circ$, the typical time for the growth of the instability is  
\begin{equation}
  \tau_{\rm RT} \approx \frac {\rho}{\sqrt{- \overrightarrow{\nabla} P \cdot \overrightarrow{\nabla} \rho}} , 
    \label{eq:RTtime} 
\end{equation}
where we assume that the wavelength of the perturbation is about the density scale height, $(d \ln \rho / dr)^{-1}$ (e.g., chapter 7 in \citealt{Priest1982}).   
 To present both stable and unstable zones in the maps to follow, we refer to the growth rate of the instability  
\begin{equation}
f_{\rm st} \equiv \frac {1}{\rho} {\sqrt{\left| \overrightarrow{\nabla} P \cdot \overrightarrow{\nabla} \rho \right|}} ~ \text{sgn} (\overrightarrow{\nabla} P \cdot \overrightarrow{\nabla} \rho ),
    \label{eq:Tst}
\end{equation}
which is a frequency in stable zones (hence the letter `f' and the subscript `st'). 
In unstable zones $f_{\rm st}<0$, and $-1/f_{\rm st}$ is approximately the growth time of the RTI (equation \ref{eq:RTtime}). In stable zones $f_{\rm st}$ is approximately the Brunt–V\"ais\"al\"a frequency (e.g., chapter 4 in \citealt{Priest1982}), assuming the density gradient is much steeper than the pressure gradient, $\vert d \ln \rho /dr \vert \gg \vert d \ln P/dr \vert$; the latter holds in our case where the low-density zone are much hotter.

\begin{figure} 
\centering
\includegraphics[width=0.42\textwidth]{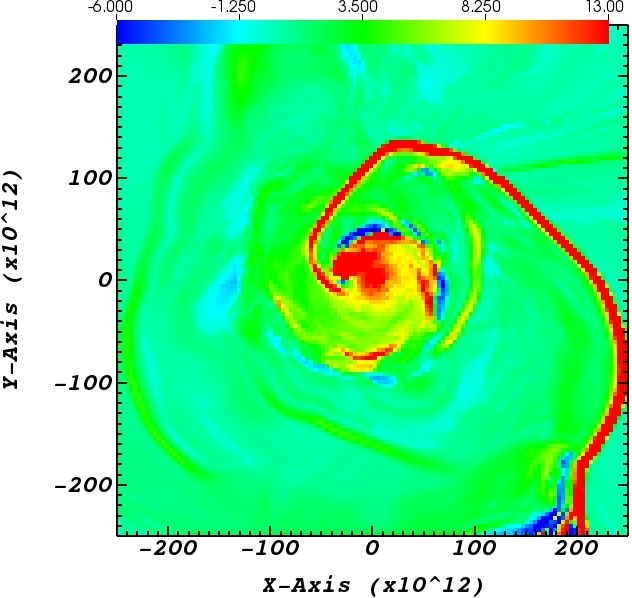}\\
\includegraphics[width=0.42\textwidth]{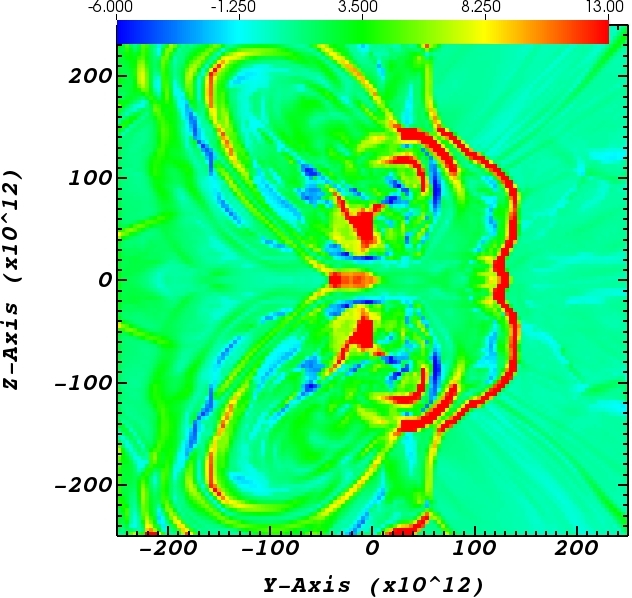}\\
\includegraphics[width=0.42\textwidth]{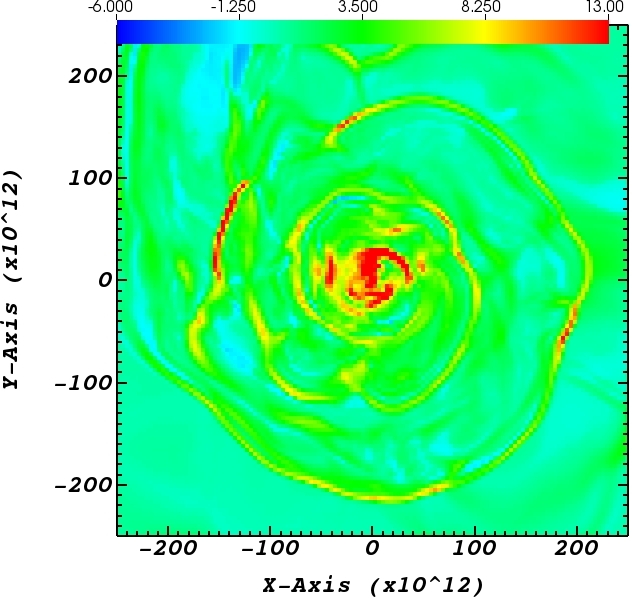}
\caption{Maps of $f_{\rm ST}$ according to equation \ref{eq:Tst}, in the $z=21 \times 10^{12} \cm$ plane at $t=2.3 \yr$ (top), $x=67 \times 10^{12} \cm$ plane at $t=2.3 \yr$ (middle), and $z=21 \times 10^{12} \cm$ plane at $t=3.8 \yr$ (bottom). Units are in $1/\yr$ according to the color bar. 
Zones with positive values are stable, and $f_{\rm ST}$ is about the Brunt–V\"ais\"al\"a frequency. Zones with negative values are unstable and $f_{\rm ST}$ is the growth rate, i.e., $f^{-1}_{\rm ST}$ is the growth time of the instability.  
}
\label{fig:RT_instability}
\end{figure}

Figure \ref{fig:RT_instability} presents maps of $f_{\rm st}$ in units of $\yr^{-1}$ for the $\alpha=0.5$ simulation in two planes and at two times. The deepest blue and pale blue zones indicate unstable zones with growth time of $\tau_{\rm RT} \simeq 0.16 \yr$ and $\tau_{\rm RT} \simeq 0.8 \yr$, respectively. As with water on oil, the unstable zones are near the interface between the two media and might have a small filling factor. Here, the Rayleigh-Taylor unstable zones with short time growth also have a small filling factor but their typical growth time is much shorter than the simulation time that we present in Figure \ref{fig:RT_instability} (2.3 and 3.8 years). Therefore, the Rayleigh-Taylor instability modes form the filaments and bubbles in the ejecta that Figure \ref{fig:DensityMaps} shows.  

The strong turbulence that the jets induce shapes the outflow and can efficiently carry energy outward, as convection plays a significant role in energy transport in CEE (e.g.,  \citealt{Sabachetal2017, Gricheneretal2018, WilsonNordhaus2019, WilsonNordhaus2020, WilsonNordhaus2022, Noughanietal2025}). 

\section{Summary} 
\label{sec:Summary}

This paper continues the 3D exploration of the role of jets that NSs launch inside the envelope of RSG stars during a CEE. The jets that the NS launches during the CEE are energetic enough to eject large amounts of mass and power a transient event that might mimic a peculiar supernova, particularly if the NS accretes from the core of the RSG star. Hence, it is termed a CEJSN. If the NS (or black hole) does not enter the core, the less energetic transient is a CEJSN impostor. 

In this study, we introduced a pre-CE envelope rotation. We found that the rotating envelope performs non-radial oscillations (Figure \ref{fig:LowHighRes} - \ref{fig:Dens_Vel_vectrQ}) and that it develops convection (Figure \ref{fig:Dens_Vel_vectrFull}). RSG stars oscillate and have extended and strong envelope convection. We, therefore, strengthen our claim from \cite{Hilleletal2023} that there is no need to relax stellar models of cool giants when transporting 1D models to a 3D numerical grid. The model oscillates and has convection, as it should. 

We do not have the computer resources to perform the simulations with jets in the high-resolution grid. The comparison of the regular and high-resolution simulations without jets (Figure \ref{fig:LowHighRes}, and \ref{fig:RotationDensity1} -  \ref{fig:Dens_Vel_vectrFull}) teaches us that the regular resolution is adequate to perform the simulations with jets, at least for our goals. 

Figures \ref{fig:DensityMaps} and \ref{fig:RT_instability} present the results of the simulations that include the jets that the NS launches inside the RSG envelope. Figure \ref{fig:DensityMaps} compares the density of the bounds and ejected gas in the non-rotating simulation with two rotating envelope simulations. The jets' power is the same in the three simulations. The most obvious difference is that the faster the rotation is, the more pronounced the spiral structure in planes parallel to the orbital plane is. The general conclusion, however, is that there are no significant differences between the simulations with different envelope rotations for the jets' power we employed. We, therefore, do not present more properties of the ejected gas, which can be found in our study of the non-rotating case in \cite{Hilleletal2023}. 

A new analysis of this study is that of Rayleigh-Taylor instabilities. Figure \ref{fig:RT_instability} presents the unstable zones in two planes and two times for the $\alpha=0.5$ simulation (equation \ref{eq:Omega} defines $\alpha$). The simulation with slower rotation and the one without envelope rotation are also prone to Rayleigh-Taylor instabilities, as the instability results from the energy deposition by jets. These instabilities account for the filaments and bubbles of the spiral structure and the ejected gas that figure \ref{fig:DensityMaps} presents.  
Clumpy ejecta (with filaments, bubbles, and clumps) influence the light curve from such events and should be the subject of future studies. 

In a broader context, jets seem to be the most prominent observable of CEE, at least in planetary nebulae with central binary stars \citep{Soker2025CEEj}. Studying the roles of jets that the secondary star launches in CEE should be a prime focus of CEE research. Our new study adds to the exploration of the role of jets in revealing some properties of Rayleigh-Taylor instabilities that jets induce.

\section*{Acknowledgments}

We thank Aldana Grichener for her valuable comments. We thank an anonymous referee for detailed comments and suggestions.  The Amnon Pazy Research Foundation supported this research.



\label{lastpage}

\end{document}